\documentstyle{elsart}

\newcommand{\be}{\begin{equation}}
\newcommand{\ee}{\end{equation}}
\newcommand{\ba}{\begin{array}{c}}
\newcommand{\ea}{\end{array}}
\newcommand{\bqa}{\begin{eqnarray}}
\newcommand{\eqa}{\end{eqnarray}}

\begin{document}

\begin{frontmatter}
\title{Some Remarks on  the Final State Interactions in $B\to \pi K$ Decays}
\author{Zuo-Hong Li, Feng Yuan and  Hanqing Zheng\thanksref{new}}
\address{Department of Physics, Peking University, Beijing 100871,
        People's Republic of China}

\thanks[new]{e--mail: zheng@ibm320h.phy.pku.edu.cn}
\begin{abstract}
Careful discussions are made on some points which are met in studying B decay 
final state interactions, taking the $B^0\to \pi^+ K^-$ 
process as an example.
We point out that $\pi$--exchange rescatterings are not important, whereas 
for $D^*$ and
$D^{**}$ exchanges, since the $B^0\to D^+D_s^-$
decay has a large branching ratio their contributions 
may be large enough
to enhance the $B\to \pi K$  branching ratio significantly.
 However our estimates
fail to predict  a large enhancement. 
\end{abstract}

\end{frontmatter}

The importance of studying final state interactions (FSIs) in the system of
the $B$ meson hadronic decay products is well known as the FSI is of strong
interaction nature and contributes the main uncertainty in extracting the
CKM matrix elements and the information of direct CP violations from $B$
decay experiments. Even though the studying of FSI effects is very difficult
it has drawn increasing attentions in recent years. Despite of some
controversies exist in the literature we believe some qualitative and/or
semi-quantitative results can be obtained~\cite{hz,rz}. The method is based
on a Regge pole model description to the partial--wave rescattering
amplitude. The contribution from absorption effects to the low partial--wave
amplitudes is also estimated and it was pointed out that, for the
meson--meson scatterings via Reggeon exchanges the absorption effects remain
small~\cite{hz}, and absorption effects further reduce the (low
partial--wave) final state rescattering effects. The theoretical tool
suitable for studying the FSI effects are the Watson--Migdal theorem for
final state interactions and the multi-channel N/D method~\cite{hz,rz}. The
main uncertainty come from inelasticity is expected to be small as the
inelastic contribution to a given exclusive $2\to 2$ rescattering is of
non-leading order and is expected to cancel each other and the net effect
remains small~\cite{hz,rz}. Under these considerations meaningful numerical
results are obtainable and it was found that for a quantity controlled by
Regge pole exchanges, the FSI effect remains small. For example, for the
charge--exchange rescattering the typical enhancement factor is of order $%
O(\lambda ^2)$~\cite{rz} where $\lambda $ is the Wolfenstein parameter. The
strong interaction phase difference between two isospin amplitudes is also
controlled by Regge pole exchanges and be a small quantity. These results
are found to be consistent with experiments~\cite{suzuki,nardulli} involving
a $D$ meson in the final decay products.

The results mentioned above implies that a calculation based on the
approximation by neglecting the FSI effects can be a reasonably good
approximation in many situations unless in the case when bare amplitudes
which can switch to each other via final state rescatterings differ by a
large amount. In such a case the tiny FSI effects are compensated by the
huge decay amplitude from which the intermediate rescattering particles are
generated. Such a process may generate sizable CP violation effects~\cite
{kamal}, therefore a careful numerical study on the rescattering effects is
very interesting physically. In this note we investigate some subtle points
which appear in the calculation of FSI effects. The first is the
rescattering process dominated by Reggeized pion exchange, which was not
discussed as much as the spin--1 exchange processes and controversial
results exist in the literature. We think it is worthwhile to give a careful
analysis on such processes. An example for rescattering process via pion
exchange is, $B^0\to K^{*+}\rho ^{-}\to K^{+}\pi ^{-}$. the pion has a tiny
mass that the pole at $t=0.02GeV^2$ is appreciately felt in the scattering
region of $t<0$, therefore the pion exchange contribution can be important.
The $B\to K\pi $ process is recently measured by experiments~\cite{exp}
which stimulates many theoretical discussions\cite{theory}. We use Eq.~(14)
derived in Ref.~\cite{rz} for our numerical studies, which reads, 
\begin{equation}  \label{fsi}
{\bf A}_{i\to j}={\bf A}_i\{{\frac{{\rm P}}\pi }\int {\frac{{\bf \rho }%
(s^{\prime }){\bf T_{ij}}(s^{\prime })}{s^{\prime }-s}}ds^{\prime }+i{\bf %
\rho }{\bf T}_{ij}\}\ ,  \label{equ2}
\end{equation}
where ${\bf A}_{i\to j}$ is a weak decay ($B\to j$) amplitude mediated by a
non-diffractive final state rescattering ($i\to j$) and ${\bf A}_i$ is the
decay amplitude renormalized by diffractive rescatterings and may be
identified as the observed physical amplitude as long as it is not too small
in the class of decay amplitudes which can switch to each other via
(non-diffractive) final state interactions. The partial wave rescattering
amplitude is $T_{ij}=T_{K^{*+}\rho ^{-}\to K^{+}\pi ^{-}}$ in the present
case.

For the simple helicity non-flip amplitudes it is argued that the absorption
effects remain small for meson--meson rescatterings.~\cite{hz} However the
pion exchange processes of helicity--flip amplitudes is well described
phenomenologically by the Williams model~\cite{irving} in which the
absorption effects is taken into account explicitly. Without such a careful
treatment to the absorption effects one can be led to misleading conclusion
as will be shown below.

For a given ($s$--channel) helicity, the pion exchange amplitude is~\cite
{irving}, 
\begin{equation}  \label{regge}
T_{\lambda _4\lambda _2}^{\lambda _3\lambda _1}=-(-t^{\prime
}/4M^2)^{n/2}(-m_\pi ^2/4M^2)^{x/2}\beta _e^{\lambda _3\lambda _1}\beta
_{\lambda _4\lambda _2}^eR[s,\alpha _e(t)]\ ,  \label{pi}
\end{equation}
where $R[s,\alpha _e(t)]$ is the Reggeized propagator, 
\begin{equation}
R[s,\alpha _e(t)]={\frac 12}[1+(-)^{s_e}e^{-i\pi \alpha _e(t)}]\Gamma
[l_e-\alpha _e(t)](\alpha ^{\prime })^{1-l_e}(\alpha ^{\prime }s)^{\alpha
_e(t)}\ ,
\end{equation}
and $l_e=s_e=0$ and $\alpha _e(t)=\alpha ^{\prime }(t-m_\pi ^2)$ for the
pion exchange ($\alpha ^{\prime }=0.93GeV^{-2}$ is the universal slope
parameter for light hadrons). In equation~(\ref{pi}) $t^{\prime }=t-t_{min}$%
, $M$ is the nucleon mass, $n=|\lambda _3-\lambda _1-\lambda _4+\lambda _2|$
is the net helicity flip and $n+x=|\lambda _3-\lambda _1|+|\lambda
_4-\lambda _2|$. In equation~(\ref{pi}) absorption effects have been taken
into account to explain the forward spike observed in experiments~\cite
{irving}. Had we omitted the absorption effects the term $(-m_\pi
^2/4M^2)^{x/2}$ in equation~(\ref{pi}) would be replaced by $(-t^{\prime
}/4M^2)^{x/2}$. Roughly speaking, the inclusion of absorption effects is
equivalent to subtracting the $s$--wave component from the full $T$ matrix
element. In our present case of $B\to K^{*^{+}}\rho ^{-}\to K^{+}\pi ^{-}$
there are two helicity amplitudes, $T_{11}$ (=$T_{-1-1}$) and $T_{00}$, with 
$n=0,x=2$ and $n=x=0$, respectively. We use the Reggeon coupling constants
from that of Ref.~\cite{irving}, $\beta _\pi ^{\lambda _\rho =1}(\rho ^0\pi
^{-})=3.45$ and $\beta _\pi ^{\lambda _\rho =0}(\rho ^0\pi ^{-})=4.40$,
respectively. Other coupling constants are estimated by SU(3) relations.
Taking the $s$--wave projection of the $T$ matrix element and making use of
Eq.( \ref{equ2}) we get, 
\begin{equation}
R^{00}\equiv {\frac{A^{00}(B\to K^{*+}\rho ^{-}\to K^{+}\pi ^{-})}{%
A^{00}(B\to K^{*+}\rho ^{-})}}=\ (0.24+0.71i)\times 10^{-2}\ .  \label{1}
\end{equation}
The absorption effects severely reduce the $s$--wave rescattering amplitude $%
T_{11}$ as can be clearly seen from Eq.~(\ref{pi}) and the rescattering
effects is negligible. If we do not take the absorption effects into account
we would obtain a much larger enhancement factor $R^{11}$ in magnitude, $%
R^{11}=(1.67-2.02i)\times 10^{-2}$. We also estimated the rescattering
amplitude $B\to K^{*^{+}}\rho ^{-}\to K^0\pi ^0$ in which $K$--exchange also
contributes (in the $u$--channel) and found that 
\begin{equation}
R^{00}=-(0.44+1.17i)\times 10^{-2}\ ,  \label{2}
\end{equation}
and $R^{11}$ again negligible after considering the absorption effects. From
the results of Eqs.~(\ref{1}) and (\ref{2}) we predict that a final state
rescattering via $\pi $ (and $K$) exchange contributes an enhancement factor
of order of $O(\lambda ^2)$, which confirms the results given in Ref.~\cite
{rz}.

It was recently claimed~\cite{charm} that the ``charm penguin'' effects
which are neglected in the naive factorization approximation have strong
effects on $B$ decay amplitudes into light hadrons. These effects are of FSI
nature, which, in the present language, imply that there may exist strong
final state rescattering effects in, say, $B\to D^{+}D_s^{-}\to \pi
^{+}K^{-} $ processes. Motivated by this we in the following estimate the
rescattering amplitude mediated by $D^{*}$ and $D^{**}$ exchanges. Based
upon an estimation on the absorptive part of the rescattering amplitude, it
was pointed out that~\cite{hz} the $D^{*}$ and $D^{**}$ Reggeons have a very
small intercept parameter ($\alpha ^0\simeq -1$ ) and therefore their
effects are much smaller than the $\rho $ exchange contributions.\footnote{%
The suppression due to the small intercept here is slightly compensated by
the slower $t$ dependence of the Regge trajectory ($\alpha^{\prime}\simeq
0.5 $), which implies a larger $s$--wave component of the rescattering
amplitude.} To give an order of magnitude estimate we use SU(4)
approximation to fix the corresponding coupling constants. We perform the
dispersive integral in equation (\ref{fsi}) from the physical threshold $%
s_{phys}^{th}=(m_{D^{+}}+m_{D_S^{-}})^2$ to $\infty $ and obtain, 
\begin{equation}
R\equiv {\frac{A(B\to D^{+}D_s^{-}\to \pi ^{+}K^{-})}{A(B\to D^{+}D_s^{-})}}%
=\ (0.56+1.76i)\times 10^{-3}\ ,  \label{3}
\end{equation}
which is one order of magnitude smaller than a typical $\rho $ exchange
contribution ($\sim O(\lambda ^2))$. The effect of large branching ratio of $%
{\rm Br}(B\to D^{+}D_s^{-})$ comparing with ${\rm Br}(B\to \pi ^{+}K^{-})$
is greatly reduced by such a tiny rescattering effect, 
\begin{equation}
|{\frac{A(B\to D^{+}D_s^{-}\to \pi ^{+}K^{-})}{A(B\to \pi ^{+}K^{-})}}|=|R|%
\sqrt{\frac{{Br(B\to D^{+}D_s^{-}}}{Br(B\to \pi ^{+}K^{-})}}\simeq 0.1\ ,
\end{equation}
in which we take $Br(B\to \pi ^{+}K^{-})\simeq 2\times 10^{-6}$ from the
naive factorization estimate~\cite{charm}. From the above result we see
that the rescattering effects is still small, if not negligible.

To conclude we pointed out that the FSI effects due to pion exchanges are
small. For the $D^{*}$ ($D^{**}$) exchange the enhancement factor is still
not large enough to explain the discrepancy between the naive factorization
results and experiments. This may imply that the formalism we use in this
note is inadequate to reveal the charm penguin effects provide that the
naive factorization estimate is reliable. Another possibility is that one
has to sum up all the intermediate states rather than $D^{+}D_s^{-}$ only.
In such a situation our method can not handle the multi-particle
intermediate states. However we insist that our method is effective in
explaining the $2\to 2$ final state rescatterings mediated by normal Reggeon
exchanges, since the dynamics of the final state interaction without charmed
particles and the one with a single charmed particle are essentially the
same and in the later case theory and experiments are found to agree with
each other.

{\it Acknowledgment}: The work of H.Z. is supported in part by China
National Natural Science Foundations.

\end{document}